\documentclass[fleqn,twoside]{article} 
\usepackage{espcrc2} 
\usepackage{epsfig} 
 
\usepackage{graphicx} 
\usepackage[figuresright]{rotating} 
 
\newcommand{\be}{\begin{equation}}  
\newcommand{\en}{\end{equation}} 
\newcommand{\bea}{\begin{eqnarray}} 
\newcommand{\ena}{\end{eqnarray}} 
 
\newcommand{\hbo}{\hbox to 1 true cm {\hfill } }  
\newcommand{\tr}{\hbox{tr}}

\newcommand{\AmS}{{\protect\the\textfont2 
  A\kern-.1667em\lower.5ex\hbox{M}\kern-.125emS}} 
 
\title{Gauge potential singularities and the gluon condensate at finite temperatures } 
 
\author{K.~Langfeld\address[tue]{Institut f\"ur Theoretische Physik,  
        Universit\"at T\"ubingen, D-72076 T\"ubingen, Germany }\thanks{  
        talk presented by KL.},  
        E.-M.~Ilgenfritz\addressmark[tue]\address[osa]{Research Center for  
        Nuclear Physics, Osaka University, Osaka 567-0047, Japan} 
        \thanks{E.-M.I. thanks for the support by the Ministry 
        of Education, Culture and Science of Japan (Monbu-Kagaku-sho) and 
        for a grant to visit CERN.}, 
        H.~Reinhardt\addressmark[tue]\thanks{ supported  
           by DFG RE856/4-1} and 
        G.~Shin\addressmark[tue]\address[snu]{Center for Theoretical  
        Physics, Seoul National University, Seoul 151-742, Korea. } 
        \thanks{ GS thanks Dong-Pil Min and Mannque Rho who made this 
        collaboration possible.} } 
        
\begin{document} 
 
\begin{abstract} 
The continuum limit of $SU(2)$ lattice gauge theory is carefully investigated  
at zero and at finite temperatures. It is found that the continuum gauge  
field has singularities originating from center degrees of freedom  
being discovered in Landau gauge.  
Our numerical results show that the density of these singularities properly  
extrapolates to a non-vanishing continuum limit.  
The action density of the non-trivial $Z_2$ links is tentatively  
identified with the gluon condensate. We find for temperatures larger  
than the deconfinement temperature that the thermal fluctuations of the 
embedded $Z_2$ gauge theory  
result in an increase of the gluon condensate with increasing temperature.  
\vspace{1pc} 
\end{abstract} 
 
\maketitle 
 
\section{INTRODUCTION} 
 
A precise definition of a Quantum Field Theory (QFT) is provided by the  
critical limit of a lattice model.  
Thereby, the QFT is solely specified by the number of space-time  
dimensions and symmetries. Here, we will critically re-investigate the  
lattice gauge theory with Wilson action which is assumed to reduce  
to continuum $SU(2)$ Yang-Mills theory in the critical limit.  
 
It is usually assumed  
that all $SU(2)$ link variables $U_{\mu}(x)$ can be expanded  
in the vicinity of the unit element for  
sufficiently small lattice spacing $a$, i.e. 
$U_\mu (x) \; = \; \exp \{ i W_\mu(x) \, a \} $,  
such that the Wilson action density reduces to the Yang-Mills Lagrangian 
\be  
\frac{\beta}{2} \tr \{ 1 - P_{\mu \nu}[U] \}  
\, \rightarrow \, 
\frac{a^4}{2g^2} F^b_{\mu \nu}[W](x) F^b_{\mu \nu}[W] \; ,  
\label{eq:1}  
\en  
where $\beta = 4/g^2$,  
$P_{\mu \nu}[U](x)$ is the plaquette calculated in terms of the  
link elements $U_{\mu}(x)$ and $F^b_{\mu \nu}[W] $ is the usual field  
strength tensor.  
We will find that this Taylor expansion is not always justified.  
We will propose to relate the action density of the  
corresponding singularities to the gluon condensate.

\section{ THE LATTICE THEORY OF GAUGE POTENTIAL SINGULARITIES } 
 
Before we can localize the links where the above expansion eventually 
fails, we must bring the link elements as close as possible  
to the unit element by exploiting the gauge freedom  
$\Omega = \prod_{x}\omega_x$.  
We are thus led to implement an algorithm which puts the Monte Carlo 
configurations into the Landau gauge, accomplishing  
$ 
\sum_{x,\mu} \; \tr \; U^{\Omega}_{\mu}(x) \; \rightarrow \; \mathrm{max},  
$ 
which is most suitable for our purpose.  
We have used an {\it improved} simulated annealing algorithm  
in order to find the maximum of the gauge fixing  
functional (details will be presented elsewhere). 
We then decompose the appropriately gauged link elements $U^{\Omega}$ into  
a center part and a coset part  
\be 
U^{\Omega}_{\mu}(x) \; = \; Z_{\mu}(x) \;  
\exp \biggl\{ i \, A^b_{\mu}(x) t^b \, a \biggr\} \; ,  
\label{eq:3}  
\en 
with $Z_{\mu}(x) = \mathrm{sign} \, \tr \; U^{\Omega}_\mu (x)$, and  
interpret $A^b_{\mu}(x)$ as the Yang-Mills gauge field in the continuum  
limit~\cite{Langfeld:2001ym}.  
An expansion in powers of $a W_\mu $ (\ref{eq:1})  
is uniformly justified only if the embedded $Z_2$ gauge theory  
spanned by the link elements $Z_{\mu}(x)$ is trivial, i.e. if $Z_{\mu}(x)=1$,  
$\forall x,\mu $.  
For a numerical check how many links $U_{\mu}(x)$ violate this condition  
it is possible directly to use the center action as singularity counter 
\be  
c \; := \; \biggl\langle 1 \,  
- \, \frac{1}{2} \tr P_{\mu \nu}[Z](x) \biggr\rangle \; ,  
\label{eq:4}  
\en 
with the plaquettes $P_{\mu \nu}[Z](x)$ expressed in terms of the center 
projected configuration. 
For the Landau gauge, one expects that the non-trivial center elements  
$Z_{\mu}(x) =-1$ are sufficiently dilute close to the continuum limit.  
An isolated non-trivial center element generates $6$ negative plaquettes 
each contributing $2$ units to the center action in (\ref{eq:4}). 
In this dilute gas approximation, we therefore find  
$c \approx 12 \, \rho \, a^4 $, where $\rho$ is the density  
of links carrying non-trivial center elements.  
\begin{figure}[t] 
\centerline{ 
\epsfxsize=\linewidth  
\epsfbox{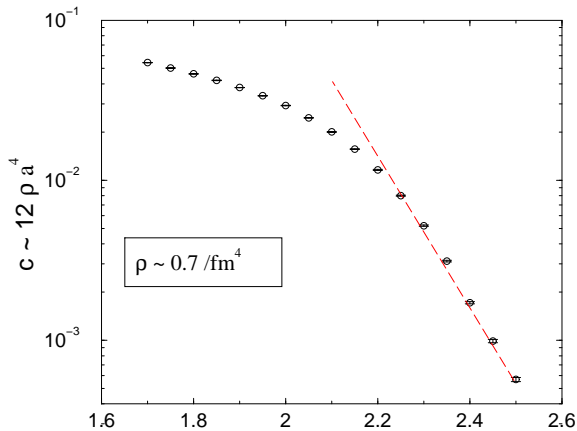} 
} 
\caption{Density of non-trivial $Z_2$ plaquettes. } 
\label{fig:1} 
\end{figure} 
Our numerical result for $c$ as function of $\beta$  
is shown in figure \ref{fig:1}. As expected, the  
singularity counter $c$ rapidly decreases with increasing $\beta $.  
Note, however, that the density $\rho$ of non-trivial center elements in  
physical units 
scales towards the continuum limit and is non-vanishing in the limiting case.

\section{ THE FINITE TEMPERATURE GLUON CONDENSATE } 
 
The Operator Product Expansion (OPE)  
somewhat artificially distinguishes between  
contributions from perturbative gluons and ''other'' contributions  
to Greenfunctions, which are not further specified.  
In lattice regularization, the perturbative gluon contribution  
is recovered by expanding the link variables (in Landau gauge)  
around the unit elements.  
Thus, the singular component, originating from the non-trivial  
center elements, contributes to the ''other parts'' which are parameterized  
by the condensates of the OPE.  
In particular, the gluon condensate is defined as the expectation  
value of the action density where the (divergent) perturbative gluon  
contribution has been subtracted.  
In practice, this subtraction uses the result of a high order calculation  
in lattice perturbation theory~\cite{bur98}. Alternatively, the gluon  
contribution can be removed by a cooling procedure which reduces the action  
of the coset (gluon) fields $A_\mu(x)$~\cite{Langfeld:2001ym}.  
The latter approach  
suggests that 
the gluon condensate $G$ gets contributions from the energy  
density stored in the underlying $Z_2$ fields, which are revealed  
by the coset cooling mechanism. We therefore find  
\be  
G \, a^4 \; \propto \; c \; = \; 12 \rho \, a^4 \; .  
\label{eq:5}  
\en 
It was already observed above (see figure \ref{fig:1}) that $G$ is  
lattice spacing independent and non-vanishing close to the continuum limit.  

\begin{figure}[t] 
\centerline{ 
\epsfxsize=\linewidth  
\epsfbox{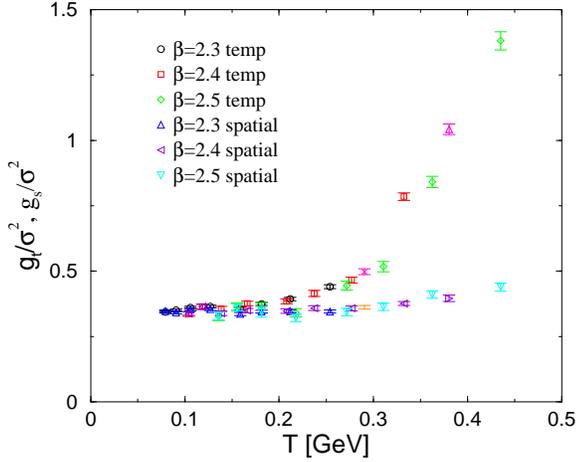} 
} 
\caption{The condensate as function of temperature. }  
\label{fig:2} 
\end{figure} 
\vskip 0.3cm  
Let us now study the temperature dependence of the gluon condensate  
$G$ which is defined by (\ref{eq:5}) and (\ref{eq:4}).  
Temperature is introduced by varying the number $N_t$ of grid  
points in time direction, $T= 1/N_t a(\beta )$. We used the one loop  
formula for $a(\beta )$ and the string tension $\sigma = (440 \,  
\mathrm{MeV} )^2 $ as the reference scale. Simulations were performed  
on $16^3 \times N_t$, $N_t=4 \ldots 16$ lattices for $\beta = 2.3, \,  
2.4, \, 2.5 $. Due to the space-time asymmetry induced by  
the finite temperature, it is convenient to independently measure  
the spatial and the time-like parts of the gluon condensate, i.e.  
$G_s$ and $G_t$. These condensates are calculated from (\ref{eq:5}) where  
$c$ (\ref{eq:4}) is separately evaluated with spatial-spatial and  
spatial-timelike plaquettes, respectively. The result is shown in figure  
\ref{fig:2}. Below the deconfinement temperature for $SU(2)$ gluondynamics,  
$T_c \, \approx \, 300 \, $MeV, 
the condensates $G_s$ and $G_t$ are weakly temperature dependent. Above $T_c$,  
the time-like component $G_t$ is rapidly increasing with $T$  
while the variations of $G_s$ with $T$ are still moderate. 
 
\vskip 0.3cm  
We interpret this behavior as follows: at zero temperature, the vacuum  
energy density of the underlying $Z_2$ gauge fields generates the  
gluon condensate. In the deconfined phase at high temperatures,  
these fields start thermally fluctuating in addition to the fluctuations of  
the coset (gluon) fields. The  
thermal energy density stored in the $Z_2$ system induces the rise  
of the gluon condensate with temperature 
while the gluonic black body radiation does not contribute by definition. 
 
\section{ DO WE HAVE TO REVISE THE CONTINUUM FORMULATION?} 
 
The ab initio continuum formulation is based on the functional integral  
over gluon fields. A choice of the  
boundary conditions for gluon fields $A^b_\mu (x)$ is included in the  
definition of the QFT. If we wish to consider singularities of the  
gauge potential, the positions of these singularities are cut out  
from the space-time manifold. This procedure generates new boundaries  
at which we must specify appropriate conditions for the gauge potential.  
Different choices of the singularities might generate theories with  
different physical contents.  
 
\vskip 0.3cm  
Let us illustrate this point in the familiar case of QED. A base manifold 
which is punctured at a closed line of space time can support a smooth  
gauge field which describes a magnetic monopole world line 
(Wu-Yang construction). Choosing several ''singular lines'' which are  
cut out from the base manifold, providing boundary conditions  
at all cuts and solving the QED partition function would result  
in a theory of photons moving in a background of given monopole  
world lines, but does not serve as a theory of monopoles.  
By contrast, the continuum limit of lattice compact QED  
represents a complete theory of photons and monopoles~\cite{Jersak:1996mn}.  
 
\section{CONCLUSIONS} 
 
Landau gauge fixing of $SU(2)$ lattice gauge theory reveals a $Z_2$ vacuum  
texture. In the continuum limit, this texture  
consists of point-like gauge potential singularities with a density of  
$\rho \approx 0.7 / fm^4$.  
The action density which is stored in the  
embedded $Z_2$ gauge system is identified with (part of) the gluon condensate.  
At temperatures  
above the deconfinement transition we observed that the $Z_2$ system carries  
thermal energy density leading to a gluon condensate which increases with  
temperature.

\end{document}